\RequirePackage{fix-cm}
\documentclass[twocolumn,showpacs,preprintnumbers,amsmath,amssymb]{revtex4}    
\usepackage{graphicx}
\usepackage{dcolumn}
\usepackage{bm}

\begin{document}

\preprint{}

\title{Nontrivial eigenvalues of the Liouvillian of an open quantum system}

\author{Ruri Nakano}
\email{rnakano@iis.u-tokyo.ac.jp}
 \affiliation{Department of Physics, Faculty of Science and Graduate School 
of Science, the University of Tokyo, 4-6-1 Komaba, Meguro, Tokyo 153-8505, Japan}

\author{Naomichi Hatano}%
\email{hatano@iis.u-tokyo.ac.jp}
\affiliation{Institute of Industrial Science, the University of Tokyo, 4-6-1 Komaba, Meguro, Tokyo 153-8505, Japan}%

\author{Tomio Petrosky}
\affiliation{
Center for Complex Quantum Systems, University of Texas at Austin, 
1 University Station, C1609, Austin, TX 78712, USA
}%

\date{\today}

\begin{abstract}
We present methods of finding complex eigenvalues of the Liouvillian 
of an open quantum system. The goal is to find eigenvalues that cannot
be predicted from the eigenvalues of the corresponding Hamiltonian.
Our model is a T-type quantum dot with an infinitely long lead.
We suggest the existence of the non-trivial eigenvalues of the Liouvillian in two ways:
one way is to show that the original problem reduces to 
the problem of a two-particle Hamiltonian
with a two-body interaction and the other way is to show that diagram
expansion of the Green's function has correlation between the bra state and the ket state.
We also introduce the integral equations
equivalent to the original eigenvalue problem.
\end{abstract}

\maketitle

\section{Introduction}
\label{intro}

If a system has a compact potential in an infinite volume, 
we can obtain complex eigenvalues of the Hamiltonian under the boundary conditions of outgoing 
waves only or incoming waves only \cite{kaisyaku,Landau}, or with the Feshbach 
formalism \cite{FB1,FB2}.
The eigenstates corresponding to the complex eigenvalues are not ghost states but have 
physical meaning; they describe resonance states, which play crucial roles in 
high-energy physics, and the imaginary part of the complex eigenvalues is the inverse lifetime
of the corresponding resonance state
\cite{Landau,Gamov,Sie,Pei,Cou,Ya.B,Humblet,Rose,Hum62,Hum641,Hum642,Mah,Rose2,Agu,E.B,B.Simon,Newton,Brandas,Kukulin},
though they do not have probabilistic interpretation 
because of divergence of their norms (See, however, Ref.~\cite{kaisyaku} for an 
extended probabilistic interpretation.). 

A discussion of the same sort can be applied to the Liouvillian,
which is the generator of the time evolution of density matrices
in the von Neumann equation:
\begin{align}
i\frac{\partial }{\partial t} \rho  = L \rho =  [H,\rho ]=z\rho,
\end{align}
where we use a unit with $\hbar =1$.
In the infinite system with a compact potential, we expect that the Liouvillian 
can have complex eigenvalues $z$
under certain boundary conditions, which have not been yet understood well, or with 
the Feshbach formalism for the Liouville operator.
For finite systems, all the  eigenvectors and the eigenvalues of the Liouvillian are
written in the forms $\rho_{nm}=|n\rangle \langle m|$ and $z_{nm}=E_{n}-E_{m}$,
respectively, where 
 $|n\rangle $ and $E_{n}$
are a discrete eigenstate and its eigenvalue of the corresponding Hamiltonian, respectively 
For infinite systems, on the other hand,
the discrete eigenvalues of the Liouvillian may be of two types:
one is $z_{\alpha \beta}=E_{\alpha }-E_{\beta }$, where $\alpha $ and $\beta $ are indices
of discrete eigenvalues
of the corresponding Hamiltonian including their complex eigenvalues, and
the other is the type not written in the form $E_{\alpha }-E_{\beta }$,
which we call nontrivial eigenvalues.
An example of the nontrivial eigenvalue for the one-dimensional
quantum gas interacting by the delta function potential has been discussed on the 
level of the weak coupling approximation by the one of the authors \cite{tomio}.
He showed that the nontrivial part of the eigenvalue gives a transport coefficient of the system.
In contrast, we here present 
another example without any 
approximation.

In the present research, we present methods of finding 
nontrivial eigenvalues of the Liouvillian of an
open quantum system.
The reason why we are interested in such eigenvalues is that the eigenstates corresponding 
to them may include physically relevant non-equilibrium states, and the imaginary part 
of the eigenvalue can be the inverse relaxation time.
In contrast to finite systems, infinite systems have dissipation of particles
into the infinity. Therefore, a state with relaxation can be 
an eigenstate of the Liouvillian under appropriate boundary conditions.
Its eigenvalue may be complex and the
corresponding eigenstate may be of the form 
$\sum_{\alpha ,\beta}c_{\alpha \beta }|\alpha \rangle \langle \beta|$
with some coefficients $c_{\alpha \beta}$.
We can find such a state not by analyzing the Hamiltonian,
only by analyzing the Liouvillian.


The paper is organized as follows.
In sec.~\ref{MA}, we explain a model that we treat in the present paper and
our approach.
In secs.~\ref{C12} and \ref{dia}, 
we present methods of finding nontrivial eigenvalues of the Liouvillian and
show evidence for their existence.
In sec.~\ref{C12}, we show it from the correspondence of the original 
Liouvillian and a two-particle Hamiltonian which has a two-body interaction.
In sec.~\ref{dia}, on the other hand, we show it
by indicating the operation of a Green's function on the bra state and the ket state simultaneously.
In appendix~\ref{Liou}, we introduce some properties of the Liouville space and our notation.
In appendix\ \ref{inte}, we also show that the original problem reduces to two
simultaneous integral equations.

\section{Model and Approach}
\label{MA}
\begin{figure}
  \includegraphics[width=1.0\columnwidth]{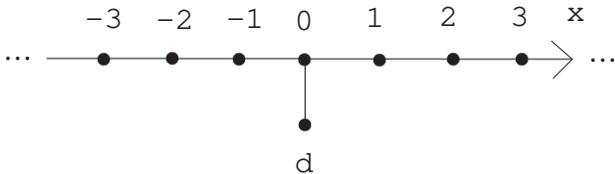}
\caption{The T-type quantum dot with an infinite lead.}
\label{fig:Model}       
\end{figure}
Let us introduce the system that we consider.
We analyze an open quantum system. The particle can then dissipate into 
the infinity.
A T-type quantum dot with an infinitely 
long lead (Fig.\ \ref{fig:Model}) is the simplest model that has the above properties.
The Hamiltonian is given by the following:
%
\begin{equation}
 H  =  -t\sum_{x=-\infty }^{\infty }(c_{x+1}^{\dagger }c_{x}+c_{x}^{\dagger }c_{x+1})
 	+(-t_{1})(d^{\dagger }c_{0}+c_{0}^{\dagger }d),
\end{equation}
where $c_{x}^{\dagger}$ and $c_{x}$ are the creation and annihilation operators at the 
site $x$ on the lead, while $d^{\dagger}$ and $d$ are those at the dot site.
We can rigorously obtain the spectrum of $H$.
In particular, the Hamiltonian has complex eigenvalues under the boundary
conditions of out-going waves only or in-coming waves only.
For this system, 
we define the projection operators $P_{\textrm{s}}$ and $Q_\textrm{s}$ as follows:
\begin{align}
P_{\textrm{s}}& = |d\rangle \langle d|,  \\
Q_{\textrm{s}} & = \sum_{x=-\infty }^{\infty }|x\rangle \langle x|,
\end{align}
and thereby we have
\begin{align}
P_{\textrm{s}}HP_{\textrm{s}} & = 0, \label{eq:PHP} \\
P_{\textrm{s}}HQ_{\textrm{s}} & = -t_{1}d^{\dagger }c_{0}, \\
Q_{\textrm{s}}HP_{\textrm{s}} & = -t_{1}c_{0}^{\dagger }d, \\
Q_{\textrm{s}}HQ_{\textrm{s}} & = 
-t\sum_{x=-\infty }^{\infty }(c_{x+1}^{\dagger }c_{x}+c_{x}^{\dagger }c_{x+1}). 
\end{align}

Our aim is to solve the eigenvalue problem of
the corresponding Liouvillian:
\begin{eqnarray}
L \rho = [H,\rho ]=z \rho \label{eq:L}
\end{eqnarray}
under appropriate boundary conditions for obtaining complex eigenvalues. 
We employ the Feshbach formalism \cite{FB1,FB2}, which 
was originally developed for the eigenvalue problem of Hamiltonians.
Then, Eq.~(\ref{eq:L}) is written in the form 
\begin{align}
&L_{\rm{eff}}(P\rho ) =  z(P\rho ),  \label{eq:Leff1}
\end{align}
where
\begin{align}
&L_{\rm{eff}}  =  PLP+PLQ\frac{1}{z-QLQ}QLP \label{eq:Leff}
\end{align}
with $P$ and $Q$ defined in Appendix \ref{Liou}.
We have
\begin{align}
PLP &=(P_{\textrm{s}}\times P_{\textrm{s}})(H\times 1-1\times H)(P_{\textrm{s}}\times P_{\textrm{s}})  \nonumber \\
 & =P_{\textrm{s}}HP_{\textrm{s}}\times P_{\textrm{s}}-P_{\textrm{s}}\times 
P_{\textrm{s}}HP_{\textrm{s}}=0
\end{align}
because of Eq.~(\ref{eq:PHP}), where the operation `$\times$' is defined in Appendix \ref{Liou}. 

Since Eq.~(\ref{eq:Leff1}) is given in the $P$ space, we need 
the element $\langle \! \langle d,d|L_{\rm{eff}}|d,d
\rangle \! \rangle $:
\begin{eqnarray}
\langle \! \langle d,d|L_{\rm{eff}}|d,d\rangle \! \rangle& = & t_{1}^{2}\langle \! \langle d,0|\frac{1}{z-QLQ}
|d,0\rangle \! \rangle\nonumber \\
& & -t_{1}^{2}\langle \! \langle d,0|\frac{1}{z-QLQ}|0,d\rangle \! \rangle  \nonumber \\
& & -t_{1}^{2}\langle \! \langle 0,d|\frac{1}{z-QLQ}|d,0\rangle \! \rangle \nonumber \\
& & +t_{1}^{2}\langle \! \langle 0,d|\frac{1}{z-QLQ}|0,d\rangle \! \rangle,\label{eq:elQLQ}
\end{eqnarray}
where 
\begin{align}
|d,d\rangle \! \rangle& :=|d\rangle \langle d|=d^{\dagger}|\textrm{vac}\rangle 
					\langle \textrm{vac}|d,  \\
|d,0\rangle \! \rangle& :=|d\rangle \langle 0|=d^{\dagger }|\textrm{vac}\rangle 
					\langle \textrm{vac}|c_{0},
\end{align}
and so on, with $|\textrm{vac}\rangle$ being the vacuum state.
Thus, 
the problem of analyzing Eq.~(\ref{eq:Leff})
is reduced to obtaining the Green's function of $QLQ$:
\begin{eqnarray}
QLQ & = & Q_{\textrm{s}}HQ_{\textrm{s}}\times 1-1\times Q_{\textrm{s}}HQ_{\textrm{s}} 
\nonumber \\
 & &  +Q_{\textrm{s}}HP_{\textrm{s}}\times Q_{\textrm{s}}
 -Q_{\textrm{s}}\times P_{\textrm{s}}HQ_{\textrm{s}}  
 \nonumber \\
 & & +P_{\textrm{s}}HQ_{\textrm{s}}\times Q_{\textrm{s}}
 -Q_{\textrm{s}}\times Q_{\textrm{s}}HP_{\textrm{s}}. \label{eq:d_QLQ}
 \end{eqnarray}

 \section{Correspondence of the one-particle Liouvillian and the two-particle Hamiltonian}
\label{C12}
Let us introduce our main result of this paper in this section.
We show that the one-particle Green's function of $QLQ$ is equal to the
two-particle Green's function
of a new Hamiltonian $\mathcal{H}$.
In order to map the eigenvalue problem of the Liouvillian to the eigenvalue problem of 
the Hamiltonian, we use some powerful methods; i.e. the Dyson equation,
closure, and so on.

Though $QLQ$ is a super-operator which acts on density operators, the density operators
belong to a Hilbert space with the inner product (\ref{eq:inpro}) in Appendix \ref{Liou}. 
Therefore, we can treat $QLQ$ as
a usual operator which acts on the Hilbert space spanned by the states $|i,j\rangle \! \rangle$.

Let us here introduce a two-particle Hamiltonian:
\begin{eqnarray}
\mathcal{H} & = & \mathcal{H}_{\textrm{lead}}^{a}+\mathcal{H}_{\textrm{lead}}^{b}+
			\mathcal{H}_{\textrm{int}} ,\label{eq:Hami} \\
\mathcal{H}_{\textrm{lead}}^{a}& = & -t\sum_{x=-\infty }^{\infty }
	(a_{x}^{\dagger }a_{x+1}+a_{x+1}^{\dagger }a_{x}) , \\
\mathcal{H}_{\textrm{lead}}^{b}& = & +t\sum_{x=-\infty }^{\infty }
	(b_{x}^{\dagger }b_{x+1}+b_{x+1}^{\dagger }b_{x}) , \\
\mathcal{H}_{\textrm{int}} & = & -t_{1}(a_{0}^{\dagger }a_{d}+a_{d}^{\dagger }a_{0})
						(1-b_{d}^{\dagger }b_{d})   \nonumber \\
					& & +t_{1}	(b_{0}^{\dagger }b_{d}+b_{d}^{\dagger }b_{0})
						(1-a_{d}^{\dagger }a_{d}),  
\end{eqnarray}
where the $a$ particle and the $b$ particle are different and mutually commutative.
Then the algebraic structure of $\mathcal{H}$ is exactly the same as that of $QLQ$
in Eq.~(\ref{eq:d_QLQ}):
\begin{align}
Q_{\textrm{s}}HQ_{\textrm{s}}\times 1 & \longleftrightarrow \mathcal{H}_{\textrm{lead}}^{a}  
\nonumber \\
-1 \times Q_{\textrm{s}}HQ_{\textrm{s}} &\longleftrightarrow \mathcal{H}_{\textrm{lead}} ^{b} 
\nonumber \\
Q_{\textrm{s}}HP_{\textrm{s}} \times Q_{\textrm{s}} &\longleftrightarrow 
			-t_{1}a_{0}^{\dagger}a_{d}(1-b_{d}^{\dagger}b_{d})   
			\nonumber \\
-Q_{\textrm{s}} \times P_{\textrm{s}}HQ_{\textrm{s}} &\longleftrightarrow 
			+t_{1}b_{0}^{\dagger}b_{d}(1-a_{d}^{\dagger}a_{d})   
			\nonumber \\
P_{\textrm{s}}HQ_{\textrm{s}} \times Q_{\textrm{s}}  &\longleftrightarrow 
			-t_{1}a_{d}^{\dagger}a_{0}(1-b_{d}^{\dagger}b_{d})  
			\nonumber  \\
-Q_{\textrm{s}} \times Q_{\textrm{s}}HP_{\textrm{s}}  & \longleftrightarrow
			+t_{1}b_{d}^{\dagger}b_{0}(1-a_{d}^{\dagger }a_{d}) \nonumber 
\end{align}
In this way, the original problem of obtaining the Green's function of $QLQ$
reduces to obtaining the Green's function of the two-particle Hamiltonian $\mathcal{H}$. 

Using the Dyson equation, we can rigorously show that
the elements of the Green's function of $\mathcal{H}$ are represented
by the Green's function of  $\mathcal{H}_{0}$, where we rearrange the Hamiltonian (\ref{eq:Hami})
as follows:
\begin{eqnarray}
\mathcal{H} & = & \mathcal{H}_{0}+\mathcal{H}_{1}=
				\mathcal{H}_{0}^{a}+\mathcal{H}_{0}^{b}+\mathcal{H}_{1}, \\
\mathcal{H}_{0}^{a}& = & -t\sum_{x=-\infty }^{\infty }
	(a_{x}^{\dagger }a_{x+1}+a_{x+1}^{\dagger }a_{x})  \nonumber \\
				& & -t_{1}(a_{d}^{\dagger}a_{0}+a_{0}^{\dagger}a_{d}) ,  \label{eq:H0a} \\
\mathcal{H}_{0}^{b}& = & +t\sum_{x=-\infty }^{\infty }
	(b_{x}^{\dagger }b_{x+1}+b_{x+1}^{\dagger }b_{x}) \nonumber \\
		& & +t_{1}(b_{d}^{\dagger }b_{0}+b_{0}^{\dagger}b_{d}), \label{eq:H0b} \\
\mathcal{H}_{1} & = & 
+t_{1}(a_{0}^{\dagger }a_{d}+a_{d}^{\dagger }a_{0})b_{d}^{\dagger }b_{d}  
	\nonumber \\
& & -t_{1}(b_{0}^{\dagger}b_{d}+b_{d}^{\dagger }b_{0})a_{d}^{\dagger }a_{d},
\end{eqnarray}
The two-particle Green's function of $\mathcal{H}_{0}$ is then given by
the convolution of one-particle Green's functions as follows:
\begin{eqnarray}
& & \langle i',j'|\frac{1}{z-\mathcal{H}_{0}}|i,j\rangle  =  \nonumber \\
	& & \frac{1}{2\pi i} 
	  \int_{-\infty }^{\infty }dE_{1}\langle i'|\frac{1}{z-E_{1}-\mathcal{H}_{0}^{a}}|i\rangle
	\langle j'|\frac{1}{E_{1}-\mathcal{H}_{0}^{b}}|j\rangle. \label{eq:convo}
\end{eqnarray}
The method of obtaining the one-particle Green's functions
 $\langle i'|(z-E_{1}-\mathcal{H}_{0}^{a})^{-1}|i\rangle$ and 
 $\langle j'|(E_{1}-\mathcal{H}_{0}^{b})^{-1}|j\rangle$ is shown in Ref.~\cite{Sasada}.

We stress here that $\mathcal{H}$ has a
two-body interaction $\mathcal{H}_{1}$.
This means that the evolution of the bra state and the ket state may correlate in
the Liouville space, which suggests that some of the 
eigenvectors are not of the form $|\alpha \rangle \langle \beta |$ and their
eigenvalues $z$ are not of
the type of $E_{\alpha }-E_{\beta }$.

\section{Diagram Expansion of the Green's Function}
\label{dia}
In the present section, we show another sign of the existence of the 
nontrivial eigenvalues, which is not the
type of $E_{\alpha }-E_{\beta }$, by describing the diagram expansion of the Green's function
of $QLQ$.

We now divide the Green's function of $QLQ$ into 
the following two parts:
\begin{align}
 QLQ =& L_{0}+L_{1},  \\
 L_{0}=&Q_{\textrm{s}}HQ_{\textrm{s}}\times 1-1\times Q_{\textrm{s}}HQ_{\textrm{s}},  \\
 L_{1}= &Q_{\textrm{s}}HP_{\textrm{s}}\times Q_{\textrm{s}}-Q_{\textrm{s}}
 \times P_{\textrm{s}}HQ_{\textrm{s}}  
 \nonumber \\
 &  +P_{s}HQ_{s}\times Q_{s}-Q_{s}\times Q_{s}HP_{s}.
\end{align}
We then define $G$ and $G_{0}$ as follows:
\begin{eqnarray}
& G  = {\displaystyle \frac{1}{z-QLQ}},  \\
& G_{0} = {\displaystyle \frac{1}{z-L_{0}}}.
\end{eqnarray}
Then we obtain
\begin{eqnarray}
G & =& G_{0} \mathop{\sum_{n=0}^{\infty }}(L_{1}G_{0})^{n}, \label{eq:G}
\end{eqnarray}
using the resolvent expansion. 

\begin{figure}
  \includegraphics[width=1.0\columnwidth]{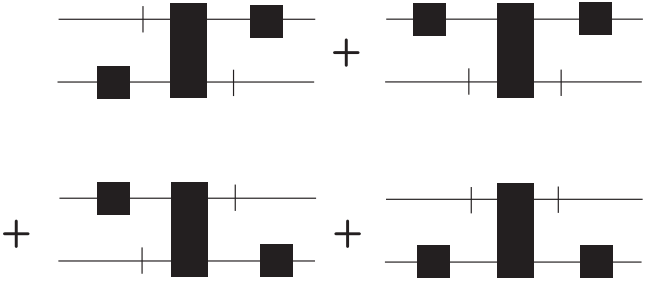}
\caption{The diagrams for the term $n=2$ of the expansion (\ref{eq:G}).}
\label{fig:n2}      
\end{figure}

Hereafter, we introduce our diagram exemplified in Fig.~\ref{fig:n2}:
\begin{itemize}
\item The upper line shows the time evolution of the bra state.
\item The lower line shows the time evolution of the ket state.
\item A thin vertical line indicates the action of $L_{1}$,which move the 
particle
from the lead to the dot or from the dot to the lead by one step.
\item A square indicates the action of $G_{0}$.
\end{itemize}
Considering that the necessary elements of $G$
are those in Eq.~(\ref{eq:elQLQ}),   
we can describe the term $n=2$ of the expansion (\ref{eq:G}) as Fig.~\ref{fig:n2}.
(The diagrams expressed in the form of integral equations are given in Appendix \ref{inte}.)

The action of $G_{0}$, which
has information that the system has an infinite volume, 
acts on the bra state and the ket state
simultaneously. This again suggests the existence of the 
nontrivial eigenvalues which cannot be
described as $E_{\alpha }-E_{\beta }$.

\section{Summary}
\label{sum}
In the present paper, we considered the eigenvalue problem of the Liouvillian of
a T-type quantum dot with an infinitely long lead and presented methods of finding nontrivial 
eigenvalues.

In secs.~\ref{C12} and \ref{dia}, 
we showed evidence for the existence of the non-trivial eigenvalues
of the Liouvillian. 
In sec.~\ref{C12}, we showed it from the correspondence of the original 
Liouvillian and the  two-particle Hamiltonian which has a two-body interaction.
In sec.~\ref{dia}, on the other hand, we showed it
by indicating the simultaneous operation
of $G_{0}$ on the bra state and ket state.
Using the result in sec.\ \ref{C12}, we have obtained nontrivial eigenvalues
approximately by discretizing the integration in Eq.~(\ref{eq:convo}), 
which may be reported elsewhere.

\appendix

\section{The Liouville space}
\label{Liou}

In the present Appendix, we introduce some properties and notations for
simplicity of calculations.
In the Liouville space, the operators are cumbersome to treat. 
The following notations are convenient for calculations.
 
 The Liouville space consists 
 of density matrices $\rho $ and is a Hilbert space with an inner product 
  $\langle \! \langle \alpha ' \beta '|\alpha \beta \rangle \! \rangle$ defined as follows;
\begin{eqnarray}
|\alpha ,\beta \rangle \! \rangle& := &  |\alpha \rangle \langle \beta |, 
 \ \ \{ |\alpha \rangle \} \ : \ \rm{CONS}, \\
\rho &=& \sum_{\alpha ,\beta }c_{\alpha \beta }|\alpha , \beta \rangle \! \rangle, 
	 \\
L|\alpha ,\beta \rangle \! \rangle& =  & [H,|\alpha \rangle \langle \beta |]
=(E_{\alpha }-E_{\beta })|\alpha \rangle \langle \beta |  \nonumber \\
& = &  (E_{\alpha }-E_{\beta })
|\alpha ,\beta \rangle \! \rangle ,\\
\langle \! \langle \alpha ',\beta '|\alpha ,\beta \rangle \! \rangle& :=  &
\mathop{\textrm{tr}}[(|\alpha '\rangle \langle \beta '|)^{\dagger }(|\alpha \rangle \langle \beta |)] = \delta_{\alpha \alpha '}\delta_{\beta '\beta }  ,   \label{eq:inpro} \\
1 & = &   \sum_{\alpha ,\beta }|\alpha ,\beta \rangle \! \rangle\langle \! \langle \alpha ,\beta |.
\end{eqnarray}
Throughout the paper, the operators that are denoted by capital letters without subscripts are in the 
Liouville space 
and operate on density matrices, while those with the subscript `$\textrm{s}$' operate on state vectors of the original Hilbert space. 
For simplification, we introduce the operation `$\times$' as follows: 
\begin{eqnarray}
(A_{\textrm{s}}\times B_{\textrm{s}})\rho & := & A_{s}\rho B_{\textrm{s}}, \\
A_{\textrm{s}}\times (bB_{\textrm{s}}+cC_{\textrm{s}}) & = & 
bA_{\textrm{s}}\times B_{\textrm{s}}+cA_{\textrm{s}}\times C_{\textrm{s}}, \\
(bB_{\textrm{s}}+cC_{\textrm{s}})\times 
A_{\textrm{s}} & = & bB_{\textrm{s}}\times A_{\textrm{s}}+cC_{\textrm{s}}\times 
A_{\textrm{s}}, \\
(A_{\textrm{s}}\times 0)\rho& = & 0 =(0\times B_{\textrm{s}})\rho , \\
(C_{\textrm{s}}\times D_{\textrm{s}})(A_{\textrm{s}}\times B_{\textrm{s}}) 
& = & C_{\textrm{s}}A_{\textrm{s}}\times B_{\textrm{s}}D_{\textrm{s}}, 
\end{eqnarray}
where $\rho$ is a density matrix and $b$ and $c$ are c-numbers. 
Then the Liouvillian is given by
\begin{eqnarray}
L & = & H\times 1-1\times H,
\end{eqnarray}
because then we have 
\begin{eqnarray}
L\rho & = & H \rho -\rho H=[H,\rho ].
\end{eqnarray}
We also define the projection 
operators
\begin{eqnarray}
P & := & P_{\textrm{s}}\times P_{\textrm{s}} , \\
Q & := & Q_{\textrm{s}}\times Q_{\textrm{s}}+P_{\textrm{s}}
\times Q_{\textrm{s}}+Q_{\textrm{s}} \times P_{\textrm{s}}, \\
P &+& Q  =  I,  \\
P_{\textrm{s}} &+& Q_{\textrm{s}}  =  I_{\textrm{s}},
\end{eqnarray}
where operators denoted by $I$ and $I_{\textrm{s}}$
are the identity operators in the respective spaces, $P$ and 
$P_{\textrm{s}}$ are 
projections on the main system, and $Q$ and $Q_{\textrm{s}}$ are projections on the environment. 

\section{Integral equations}
\label{inte}
In the present Appendix, we show that obtaining the elements of the Green's function of $QLQ$
is equivalent to solving two simultaneous integral equations.

Using the Dyson equation up to the second order,
\begin{align}
G=G_{0}+G_{0}L_{1}G_{0}L_{1}G,
\end{align}
where the first-order term vanishes,
we obtain the relations:
\begin{align}
\langle \! \langle x,d|G|0,d\rangle \! \rangle
&= 
\langle \! \langle x,d|G_{0}|0,d \rangle \! \rangle \nonumber \\
&-t_{1}^{2}\sum_{x_{1},x_{3}}\langle \! \langle x,d|G_{0}|x_{1},d\rangle \! \rangle
	\nonumber \\
 	&\langle \! \langle x_{1},0|G_{0}|0,x_{3}\rangle \! \rangle
	\langle \! \langle d,x_{3}|G|0,d\rangle \! \rangle \nonumber \\
	&+t_{1}^{2}\sum_{x_{1},x_{2}}\langle \! \langle x,d|G_{0}|x_{1},d\rangle \! \rangle
	\nonumber \\
	&\langle \! \langle x_{1},0|G_{0}|x_{2},0\rangle \! \rangle
	\langle \! \langle x_{2},d|G|0,d\rangle \! \rangle, \label{eq:xdG0d}  \\
\langle \! \langle d,x|G|0,d\rangle \! \rangle
&= 
t_{1}^{2}\sum_{x_{1},x_{3}}\langle \! \langle d,x|G_{0}|d,x_{1}\rangle \! \rangle
	\nonumber \\
 	&\langle \! \langle 0,x_{1}|G_{0}|0,x_{3}\rangle \! \rangle
	\langle \! \langle d,x_{3}|G|0,d\rangle \! \rangle \nonumber \\
	&-t_{1}^{2}\sum_{x_{1},x_{2}}\langle \! \langle d,x|G_{0}|d,x_{1}\rangle \! \rangle
	\nonumber \\
	&\langle \! \langle 0,x_{1}|G_{0}|x_{2},0\rangle \! \rangle
	\langle \! \langle x_{2},d|G|0,d\rangle \! \rangle, \label{eq:dxG0d}  
\end{align}
\begin{align}
\langle \! \langle d,x|G|d,0\rangle \! \rangle
&= 
\langle \! \langle d,x|G_{0}|d,0 \rangle \! \rangle \nonumber \\
&+t_{1}^{2}\sum_{x_{1},x_{3}}\langle \! \langle d,x|G_{0}|d,x_{1}\rangle \! \rangle
	\nonumber \\
 	&\langle \! \langle 0,x_{1}|G_{0}|0,x_{3}\rangle \! \rangle
	\langle \! \langle d,x_{3}|G|d,0\rangle \! \rangle \nonumber \\
	&-t_{1}^{2}\sum_{x_{1},x_{2}}\langle \! \langle d,x|G_{0}|d,x_{1}\rangle \! \rangle
	\nonumber \\
	&\langle \! \langle 0,x_{1}|G_{0}|x_{2},0\rangle \! \rangle
	\langle \! \langle x_{2},d|G|d,0\rangle \! \rangle, \label{eq:dxGd0}  \\
\langle \! \langle x,d|G|d,0\rangle \! \rangle
&= 
-t_{1}^{2}\sum_{x_{1},x_{3}}\langle \! \langle x,d|G_{0}|x_{1},d \rangle \! \rangle
	\nonumber \\
 	&\langle \! \langle x_{1},0|G_{0}|0,x_{3}\rangle \! \rangle
	\langle \! \langle d,x_{3}|G|d,0\rangle \! \rangle \nonumber \\
	&+t_{1}^{2}\sum_{x_{1},x_{2}}\langle \! \langle x,d|G_{0}|x_{1},d\rangle \! \rangle
	\nonumber \\
	&\langle \! \langle x_{1},0|G_{0}|x_{2},0\rangle \! \rangle
	\langle \! \langle x_{2},d|G|d,0\rangle \! \rangle. \label{eq:xdGd0}  
\end{align}
We define the Fourier elements of $\langle \! \langle x,d|G|0,d\rangle \! \rangle$,
$\langle \! \langle d,x|G|0,d\rangle \! \rangle$, $\langle \! \langle d,x|G|d,0\rangle \! \rangle$,
and $\langle \! \langle x,d|G|d,0\rangle \! \rangle$
as $\mathop{G^{(0d)}}(k;d)$, $\mathop{G^{(0d)}}(d;k)$, $\mathop{G^{(d0)}}(d;k)$ and
$\mathop{G^{(d0)}}(k;d)$. Then we obtain two simultaneous integral
equations:
\begin{widetext}
\begin{eqnarray}
\left \{
\begin{array}{rcl}
 \mathop{G^{(0d)}}(k;d) 
	&=& \frac{1}{z+2t\cos k}  
	 \quad -t_{1}^{2}\int_{-\pi }^{\pi }\frac{dk_{2}}{2\pi }
	\frac{\mathop{G^{(0d)}}(d;k)}{(z+2t \cos{k})(z+2t\cos k-2t\cos k_{2})}  
	  +t_{1}^{2} \mathop{G^{(0d)}(k;d)}\int_{-\pi }^{\pi }\frac{dk_{2}}{2\pi }
	\frac{1}{(z+2t\cos k)(z+2t\cos k-2t\cos k_{2})} , \\
 \mathop{G^{(0d)}}(d;k) 
 	& = &  t_{1}^{2}\mathop{G^{(0d)}}(d;k)
	\int_{-\pi }^{\pi }\frac{dk_{1}}{2\pi }\frac{1}{(z-2t\cos{k})(z+2t\cos{k_{1}}-2t\cos{k})}
	\quad  -t_{1}^{2}\int_{-\pi }^{\pi }\frac{dk_{1}}{2\pi }
	\frac{\mathop{G^{(0d)}(k_{1};d)}}{(z-2t\cos{k})(z+2t\cos_{k_{1}}-2t\cos{k})} ,  \\
\end{array}
\right.
\end{eqnarray}
\begin{eqnarray}
\left \{
\begin{array}{rcl}
 \mathop{G^{(d0)}}(d;k) & = & \frac{1}{z-2t\cos k}   
	- t_{1}^{2}\int_{-\pi }^{\pi }\frac{dk_{2}}{2\pi }
	\frac{\mathop{G^{(d0)}}(k_{2};d)}{(z-2t \cos{k})(z-2t\cos k+2t\cos k_{2})}  
	 +t_{1}^{2}\mathop{G^{(d0)}(d;k)}\int_{-\pi }^{\pi }\frac{dk_{2}}{2\pi }
      \frac{1}{(z-2t\cos k)(z-2t\cos k+2t\cos k_{2})} ,\\
 \mathop{G^{(d0)}}(k;d) &=& 
 	\quad t_{1}^{2}\mathop{G^{(d0)}}(k;d)
	\int_{-\pi }^{\pi }\frac{dk_{1}}{2\pi }\frac{1}{(z+2t\cos{k})(z-2t\cos{k_{1}}+2t\cos{k})}
 	\quad -t_{1}^{2}\int_{-\pi }^{\pi }\frac{dk_{2}}{2\pi }
	\frac{\mathop{G^{(d0)}(d;k_{2})}}{(z+2t\cos{k})(z-2t\cos_{k_{2}}+2t\cos{k})} .  
\end{array}
\right.
\end{eqnarray}
\end{widetext}
Then, the original problem reduces to solving the above two simultaneous integral
equations.
In other words, solving the problem in the ways of secs.~\ref{C12} and \ref{dia} 
gives a solution of the integral equations.

%
%




\end{document}